\documentclass[aps]{revtex4}

\def\.{\!\cdot\!}
\def\:{\cdots}
\def\[{\left[}
\def\]{\right]}
\def\({\left(}
\def\){\right)}

\def\bi{\begin{itemize}}

\def\ei{\end{itemize}}
\def\be{\begin{eqnarray}}
\def\ee{\end{eqnarray}}
\def\bn{\begin{enumerate}}
\def\en{\end{enumerate}}
\def\h{{1\over 2}}

\def\nn{\nonumber}

\def\r2{\sqrt{2}}

\def\L{{\cal L}}

\def\rr2{{1\over\sqrt{2}}}

\begin{document}
\title{What if $\theta_{13}$ Is Small?}
\author{C.S. Lam}
\address{Department of Physics, McGill University, Montreal, QC, Canada H3A 2T8\\ and\\
Department of Physics and Astronomy, University of British Columbia,  Vancouver, BC, Canada V6T 1Z1 \\
\email: Lam@physics.mcgill.ca}

\begin{abstract}

In the basis where the charged lepton mass matrix is diagonal,
the left-handed neutrino mass matrix is invariant under the permutation of the second and third generations if, and only if, the reactor angle $\theta_{13}$ is zero and the atmospheric mixing angle $\theta_{23}$ is maximal. 
In the presence of the seesaw mechanism, this
symmetry leads to an inverted hierarchy, with $m_3=0$. This inverted mass spectrum is doubly protected
if the right-handed neutrinos also have a 2-3 symmetry.
\end{abstract}
\maketitle

This article is based on a talk given at Theory Canada I, held in Vancouver on June 2-5, 2005.

Neutrino is the most abundant particle in the universe. It is important in astrophysical and cosmological
events, and it may provide a window into physics beyond the Standard Model. 
For these reasons its property
should be better understood.

We know \cite{ALTARELLI} that the root mean squared mass gap between the second and the first neutrinos  is 
$\Delta m_\odot=\sqrt{m_2^2-m_1^2}\simeq 9$ meV, with the second heavier than the first. We also know that
the atmospheric mass gap $\Delta m_{atm}=\sqrt{|m_2^2-m_3^2|}\simeq 52$ meV, but this time we do not know
which of the two masses is heavier. As to the absolute mass scale, 
only upper bounds are known. Tritium $\beta$-decay
provides an upper bound of slightly larger than 2 eV for the electron-neutrino mass, 
and astrophysical measurements have claimed to give an upper 
bound of 0.47 eV for the sum of the three masses.

The mixing of the three left-handed neutrinos is described by three 
mixing angles, $\theta_\odot=\theta_{12},\
\theta_{atm}=\theta_{23},\ \theta_{reac}=\theta_{13}$, and a CP phase
angle $\delta$. In addition, there are
two relative phase angles for the three Majorana mass parameters, 
$\phi_{12}$ and $\phi_{13}$. None of the phases have been measured, but two of the mixing
angles are known.
The solar angle $\theta_\odot$ has a value of about 32$^\circ$, the atmospheric 
angle $\theta_{23}$
is known to be maximal at 45$^\circ$, but only an upper bound of about $12^\circ$ is known for the reactor angle
$\theta_{reac}$.

Suppose future measurement of $\theta_{reac}$ gives it a very small value, so small that we can approximate it by zero. 
It is the purpose of this talk to show that in that case,
the neutrino spectrum is inverted, and the lowest mass is zero. In other words,
$m_2>m_1>m_3=0$. Since the mass gaps are known, all masses can now be computed. For example,
the mass parameter governing the neutrinoless double $\beta$-decay is then given by
$m_{ee}\simeq (52{\rm \ meV})\sqrt{1-0.84\sin^2(\phi_{12}/2)}$.

This is an interesting but sad scenario. Such small masses
and small reactor angle make them difficult to be
measured in the near future.

The argument 
for this assertion goes in two steps \cite{LAM1,LAM2}. First, one shows that a necessary 
and sufficient condition for $\theta_{23}=45^\circ$ and
$\theta_{13}=0$ is that, in the basis where the charged lepton mass matrix is diagonal.
 the left-handed neutrino Majorana mass matrix $M_\nu$ has a 2-3 symmetry.
Namely, it is invariant under
the simultaneous permutation of the second and third rows as well as the second and third columns. 

Next, one invokes the seesaw mechanism to write $M_\nu$ in the factorized form $M_D^TM_R^{-1}M_D$, and show that
as a consequence of the 2-3 symmetry, the Dirac mass matrix $M_D$ has a zero determinant. It follows
from the seesaw formula that
the determinant of $M_\nu$ also vanishes. It is then easy to see that the resulting zero eigenvalue is $m_3$.

Now we proceed to discuss the arguments in the first step. The leptonic mass Lagrangian can be written symbolically as
\be
\L=-\bar L_e M_e R_e-L_\nu^T M_\nu L_\nu,\ee
where $L_e, L_\nu$ are the left-handed fields respectively for the charged leptons and neutrinos, and $R_e, R_\nu$ are the
corresponding right-handed fields.
$M_e$ is the $3\times 3$ charged lepton mass matrix, and $M_\nu$ is the symmetric mass matrix for 
left-handed Majorana neutrinos. 

In the basis where $M_e={\rm diag}(m_e,m_\mu,m_\tau)$ is diagonal, $M_\nu$ can be diagonalized 
by the unitary MNS mixing matrix \cite{MNS} $U$, $M_\nu=UM_\nu^dU^T$, with $M_\nu^d={\rm diag}(\tilde m_1,
\tilde m_2,\tilde m_3)$
being the (generally complex) neutrino mass parameters. The absolute values $|\tilde m_i|$ of these parameters are the 
physical masses $m_i$. Without loss of generality, we can take $\tilde m_1$ to be real, and the phases of
$\tilde m_2,\tilde m_3$ to be $\phi_{12}$ and $\phi_{13}$ respectively.

With $\theta_{23}=45^\circ$ and $\theta_{13}=0$ zero, the mixing matrix $U$ can be parameterized as
\be U={1\over \r2}\pmatrix{\r2 c&\r2 s&0\cr -s&c&1\cr -s&c&-1\cr},\ee
where $s=\sin\theta_{12}$, $c=\cos\theta_{12}$, and $\theta_{12}$ is the solar mixing angle.

The resulting neutrino mass matrix 
\be 
M_\nu=UM_\nu^dU^T=
\pmatrix{\nu_{11}&\nu_{12}&\nu_{12}\cr \nu_{12}&\nu_{22}&\nu_{23}\cr \nu_{12}&\nu_{23}&\nu_{22}\cr},\ee
with
\be
\nu_{11}&=&c^2\tilde m_1+s^2\tilde m_2\nn\\
\nu_{12}&=&cs(\tilde m_2-\tilde m_1)/\r2\nn\\
\nu_{22}&=&\h(s^2\tilde m_1+c^2\tilde m_2+\tilde m_3)\nn\\
\nu_{23}&=&\h(s^2\tilde m_1+c^2\tilde m_2-\tilde m_3),\ee
is invariant under the simultaneous interchanges of the second and third columns, 
together with the second and third rows. We shall refer to this symmetry as the 2-3 symmetry
for left-handed neutrinos.

Conversely, it can be shown that a neutrino mass matrix with a 2-3 symmetry leads to a maximal
atmospheric angle and a zero reactor angle \cite{LAM1,OTHERS}.

This symmetry can be formally written as $M_\nu=PM_\nu P$, where $P$ is the 2-3 permutation matrix
$$P=\pmatrix{1&0&0\cr 0&0&1\cr 0&1&0\cr}.$$
Inserting this relation into (1), we see that the Lagrangian is invariant under a 2-3 permutation
of the neutrino fields, $L_\nu\to PL_\nu$. The converse is also true.

We can now proceed to the second step of the argument. With the seesaw hypothesis, the neutrino
mass term in (1) is obtained from the Lagrangian $\L_\nu=-L_\nu^\dagger M_D R_\nu-\h R_\nu^T M_R R_\nu$
by eliminating the right-handed neutrino field $R_\nu$. The result is the seesaw formula
$M_\nu=M_D^TM_R^{-1}M_D$. In the presence of a 2-3 symmetry of $L_\nu$, the invariance of $\L_\nu$
implies $L_\nu^\dagger M_DR_\nu=L_\nu^\dagger PM_DR_\nu$. Hence $M_D=PM_D$, so $M_D$ is invariant
under the interchange of the second and third rows. Such a matrix must have a zero determinant. Using
the seesaw formula, $M_\nu=M_D^TM_R^{-1}M_D$ must also have a zero determinant. Using (3) and (4),
it is easy to show that the zero eigenvalue is $m_3$.

This completes the demonstration that $m_2>m_1>m_3=0$ if and only if the atmospheric angle is
maximal, and the reactor angle is zero.

If the Lagrangian $\L_\nu$ is also invariant under a 2-3 permutation of right-handed neutrino field,
$R_\nu\to PR_\nu$, then a similar argument demands that $M_D=M_DP$. This also leads to $\det M_\nu=\det M_D=0$,
and $m_3=0$. The inverted mass spectrum $m_2>m_1>m_3=0$ is then doubly protected, so that
this spectrum is expected to remain qualitatively valid even with a small breaking of the 2-3 symmetries.

This research is supported by the Natural Sciences and Engineering Research Council of Canada.


\end{document}